\documentclass[aps,prl,showpacs,twocolumn,superscriptaddress]{revtex4-2}
\usepackage{ae}
\usepackage[T1]{fontenc}
\usepackage[ansinew]{inputenc}
\usepackage{amsmath}
\usepackage{amssymb}
\usepackage{graphicx}
\usepackage{color}

\usepackage[colorlinks]{hyperref}
\usepackage{epstopdf}

\usepackage{soul,xcolor}
\usepackage{soul,xcolor}
\setstcolor{red}

\makeatletter
\@ifundefined{textcolor}{}
{%
 \definecolor{BLACK}{gray}{0}
 \definecolor{WHITE}{gray}{1}
 \definecolor{RED}{rgb}{1,0,0}
 \definecolor{GREEN}{rgb}{0,1,0}
 \definecolor{BLUE}{rgb}{0,0,1}
 \definecolor{CYAN}{cmyk}{1,0,0,0}
 \definecolor{MAGENTA}{cmyk}{0,1,0,0}
 \definecolor{YELLOW}{cmyk}{0,0,1,0}
\definecolor{ORANGE}{rgb}{1,0,1} }

\@ifundefined{definecolor}
 {\@ifundefined{definecolor}
 {\@ifundefined{definecolor}
 {\@ifundefined{definecolor}
 {\@ifundefined{definecolor}
 {\usepackage{color}}{}
}{}
}{}
}{}
}{}

\usepackage{epsfig}

\def\b0{{\bf{0}}}

\begin{document}

\title{
%Coexistence of spin and charge density wave {\color{red}phases} and  \\ {\color{red}incompressible regions with incommensurate fillings} in confined fermionic  chains}
Coexistence of insulating phases in confined \\fermionic chains with a Wannier-Stark potential}

%%%%%%%%%%%%%%%%%%%%%%%%%%%%%%%%%%%%%%%%%%%%%%%%%%%%%
%%%%%%%%%%%%%%%%%%%%%%%%%%%%%%%%%%%%% %%%%%%%%%%%%%%%

%%%%%%%%%%%%%%%%%%%%%
%%%%%%%%%%%%%%%%%%%%%%%%%%%%%%%%%%%%%%%%%%%%%%%%%%%%%
\author{N. Aucar Boidi}
\affiliation{Centro At\'{o}mico Bariloche, Instituto Balseiro, 8400 Bariloche, Argentina}
\email{nairaucar@gmail.com}

\author{K. Hallberg}
\affiliation{Centro At\'{o}mico Bariloche, Instituto Balseiro, 8400 Bariloche, Argentina}
\affiliation{Instituto de Nanociencia y Nanotecnolog\'{\i}a CNEA-CONICET, 8400 Bariloche, Argentina}
\email{karenhallberg@gmail.com}
%%%%%%%%%%%%%%%%%%%%%%%%%%%%%%%%%%%%%%%%%%%%%%%%%%%%%

\author{Amnon Aharony}
\affiliation{School of Physics and Astronomy, Tel Aviv University, Tel Aviv 6997801, Israel}
\email{aaharonyaa@gmail.com}

\author{Ora Entin-Wohlman}
\affiliation{School of Physics and Astronomy, Tel Aviv University, Tel Aviv 6997801, Israel}
\email{orawohlman@gmail.com}

\date{\today}

%%%%%%%%%%%%%%%%%%%%%%%%%%%%%%%%%%%%%%%%%%%%%%%%%%%%%
%%%%%%%%%%%%%%%%%%%%%%%%%%%%%%%%%%%%%%%%%%%%%%%%%%%%%

\begin{abstract}
\noindent

We study  fermions  on a finite chain, interacting repulsively when residing   on the same  and on nearest-neighbor sites, and subjected to a Wannier-Stark linearly-varying potential.
Using the density matrix renormalization-group numerical technique to solve this generalized extended Hubbard model,  the ground state exhibits a staircase of (quasi) plateaus in the average local site  density along the chain, decreasing from being doubly-filled to empty as the potential increases. These `plateaus' represent locked-in commensurate phases of charge density waves together with band and Mott insulators. These phases  are separated by incompressible regions with incommensurate fillings. It is suggested that experimental variations of the slope of the potential and of the range of the  repulsive interactions will produce such a coexistence of phases which have been individually expected theoretically and observed experimentally for uniform systems.
\end{abstract}

%%%%%%%%%%%%%%%%%%%%%%%%%%%%%%%%%%%%%%%%%%%%%%%%%%%%%
%%%%%%%%%%%%%%%%%%%%%%%%%%%%%%%%%%%%%%%%%%%%%%%%%%%%%

\maketitle

%%%%%%%%%%%%%%%%%%%%%%%%%%%%%%%%%%%%%%%%%%%%%%%%%%%%%
%%%%%%%%%%%%%%%%%%%%%%%%%%%%%%%%%%%%%%%%%%%%%%%%%%%%%

\noindent{\it Introduction.---} The complexity of quantum many-body systems originates from the interplay of strong
interactions, quantum statistics, and the large number of quantum-mechanical degrees of
freedom. This interplay generates a multitude of phases, e.g., insulating commensurate charge (CDW) and spin (SDW) density waves and compressible (metallic) phases. This complexity already shows up in one dimension, in which one can use (and test) a variety of theoretical and experimental tools for their study. The simplest picture for interacting particles in one dimension  (1D)  is given by  the Hubbard Hamiltonian, which includes interactions, $U$, only between particles residing on the same lattice site~\cite{hubbard}. This interaction competes with the kinetic [nearest-neighbor (nn) tunneling] energy, $t$, resulting for instance, in antiferromagnetic structures \cite{Anderson}.

However, this simple  Hamiltonian cannot reproduce certain phases, like charge  density-waves.
Those are generated {\it e.g.} by the {\it extended Hubbard model}, which also includes nn interactions, $V$. Its one-dimensional version   reveals a rich phase diagram, which includes the band and Mott insulating phase \cite{Mott},  SDW and CDW and metallic phases~\cite{Mila,Ejima,Glocke,Zhang}.  It has also been used to describe data collected in experiments performed on chains of cold atoms \cite{bloch,guan}. In higher dimensions, it has been used to describe  bulk and edge states in electronic insulators~\cite{HF}.

An exact analytic solution of the extended Hubbard Hamiltonian, in particular  on a finite chain (the system amenable to cold-atom experiments), has not yet been found. It has been studied  by  a variety of numerical and approximate methods (e.g., Refs. \cite{hirsch,Nishino,Furusaki,Spalding,phased,Philoxene}), emphasizing  the  half-filled case, where one finds (for fermions in 1D), the insulating Mott antiferromagnetic phase~\cite{Mott} and CDW phases.

Experiments  on cold-atom arrays
naturally involve finite samples.  Numerical calculations  performed on such systems used various boundary conditions:
hard walls,  periodic and open boundaries, or potentials representing confining harmonic traps \cite{guan,parabola,heidrich}.
These works concentrate mostly on the region around the `center' of the confined structure, whose details are  usually not sensitive to the particular form of the boundaries, and so  its possible structures  are determined by  $U,~V$ and particle density $n$. Remarkably, experiments (e.g., on cold atoms) have observed some of the theoretically predicted phases~\cite{gross,greiner}.
 Less  attention has been paid to the structures near the `edges' of the samples and to their dependence on the details of the boundary conditions, in particular when the confinement is achieved by varying  site energies. Such a confining scheme has been recently considered,
using the self-consistent Hartree-Fock approximation, for the two-dimensional extended Hubbard Hamiltonian,
and found coexistence of various structures (phases) near the free ends of the samples~\cite{HF}.

In this Letter we generalize the extended Hubbard Hamiltonian to a 1D fermionic chain, confined by a {\it linear  potential}, which mimics either edge configurations in bulk systems or cold-atom arrays placed in an electric field. Such a potential can be produced by a longitudinal electric field, as in the Wannier-Stark model~\cite{stark}.
HERE

Given the
complex nature of the many-body problem associated with our system, we resort to one of the most accurate numerical methods for correlated systems, the density matrix renormalization-group (DMRG) \cite{white,whiteprb,reviewkaren,Uli,Uli2,banuls}, which uses quantum information to keep the most relevant states.
As we show, the linear potential generates in the ground state the simultaneous existence of segments in which different phases coexist, each of which having been observed separately before, on long uniform chains. Our results are presented by plots of the local quantum-averaged density  on the sites $i$ on the chain,  $\langle n^{}_i\rangle$,
the nn density-density correlations $\langle n_in_{i+1}\rangle$ and the nn spin-spin correlations $\langle s^z_is^z_{i+1}\rangle$,
(e.g., Fig. \ref{static}). Instead of a smooth decrease, the local average of $\langle n^{}_i\rangle$ shows flat steps, corresponding to locked-in Mott or CDW structures (e.g., $212121\dots$, $101010\dots$, \cite{Com}). These locked-in steps are similar to those observed for commensurate wave vectors in the devil's staircase~\cite{bak,lockin}.   Between these steps,
 $\langle n^{}_i\rangle$ decreases more smoothly, representing incommensurate regions, which can be thought of as `domain walls' with varying lengths~\cite{Cho}. As shown below, the local density of states on these intermediate sites exhibits small energy gaps, which imply that they are incompressible  (insulating),  in spite of having incommensurate fillings. We will refer to them hereafter as incompressible incommensurate-filling phases `IIF'.
The specific sequence of phases, and their sizes, can be modified experimentally, e.g.,  by changing the slope of the potential.
Neighboring structures in a sequence are often also neighboring in the phase diagrams
found for uniform systems (which are not subjected to the linear potential).

%%%%%%%%%%%%%%%%%%%%%%%%%%%%%%%%%%%%%%%%%%%%%%%%%%%%%
%%%%%%%%%%%%%%%%%%%%%%%%%%%%%%%%%%%%%%%%%%%%%%%%%%%%%

\vspace{0.2cm}

%%%%%%%%%%%%%%%%%%%%%%%%%%%%%%%%%%%%%%%%%%%%%%%%%%%%%
%%%%%%%%%%%%%%%%%%%%%%%%%%%%%%%%%%%%%%%%%%%%%%%%%%%%%

\noindent{\it Model.---} We study the generalized 1D extended Hubbard Hamiltonian
\begin{align}
{\cal H}=&-t\sum_{i,\sigma} \big(c^{\dagger}_{i,\sigma}c^{}_{i+1,\sigma} +{\rm  h.c.}\big)  + \sum_i (\mu^{}_i -\mu)n^{}_i\nonumber\\
&+ U \sum_i \big(n^{}_{i,\uparrow}-1/2\big)\big(n^{}_{i,\downarrow}-1/2\big) \nonumber\\
&+ V \sum_{i} \big(n^{}_{i}-1\big)\big( n^{}_{i+1}-1\big) \ ,
\label{hamil}
\end{align}
%\begin{align}
%H=& -t\sum_{i,\sigma} (c^{\dagger}_{i,\sigma}c^{}_{i+1,\sigma} +{\rm  h.c.}) + U \sum_i n^{}_{i,\uparrow}
%n^{}_{i,\downarrow} \nonumber\\
%&+ V \sum_{i,\sigma,\sigma'} n^{}_{i,\sigma} n^{}_{i+1,\sigma'} + \sum_i \left(\mu^{}_i -\mu +\epsilon  \right)n^{}_i \ ,
%\label{hamil}
%\end{align}
where $i$ is the site index, $i=0,\dots,L-1$   (we consider an odd number of sites without loss of generality). Here, $\mu$ is the fixed external chemical potential,
$c^\dagger_{i,\sigma}$ creates an electron with spin $\sigma(=\uparrow,\downarrow)$ at site $i$, $n^{}_{i,\sigma}=c^\dagger_{i,\sigma}c^{}_{i,\sigma}$, $n^{}_i=n^{}_{i,\uparrow}+n^{}_{i,\downarrow}$, while $U$ and $V$ are the repulsive interactions  between electrons on the same  and  nn sites, respectively (see Fig. \ref{system}). The site-dependent local energy (the Wannier-Stark potential)
$\mu^{}_i$ describes a linear external  potential,
\begin{align}
\mu^{}_i=\mu^{}_0[i/i^{}_c-1]\ .
\label{mui}
\end{align}
The site $i^{}_c=(L-1)/2$ represents the center of the `edge', where $\mu^{}_{i^{}_c}=0$.
The particular form of ${\cal H}$ was chosen so that at $\mu=0$ (up to a constant energy) it is particle-hole symmetric when $i\rightarrow L-1-i$ and $n^{}_i\rightarrow 2-n^{}_i$. In that case we always have  $n^{}_{i^{}_c}=1$.

For an infinite chain, $\mu^{}_i$ is large and negative at large and negative $i$, and therefore we expect all the sites there to be filled, i.e.,$n^{}_i=n^{}_{i,\uparrow}+n^{}_{i,\downarrow}=2$. Similarly, $\mu^{}_i$ is large and positive at large and positive $i$, and therefore we expect all the sites there to be empty, i.e. $n^{}_i=0$, as drawn in Fig. \ref{system}. For a finite chain, as we use here, this is still expected for a large slope, $\mu^{}_0\gg 1$, when the whole `edge' between the fully-occupied and empty `phases' is confined within the chain. Indeed, this is confirmed by our calculations. However,
the `end' trivial phases disappear for small slopes, for which the observed structures depend on the open boundaries.

%%%%%%%%%%%%%%%%%%%%%%%%%%%%%%%%%%%%%%%%%%%%%%%%%%%%%
%%%%%%%%%%%%%%%%%%%%%%%%%%%%%%%%%%%%%%%%%%%%%%%%%%%%%

\begin{figure}[h!]
\includegraphics[width=0.5\textwidth]{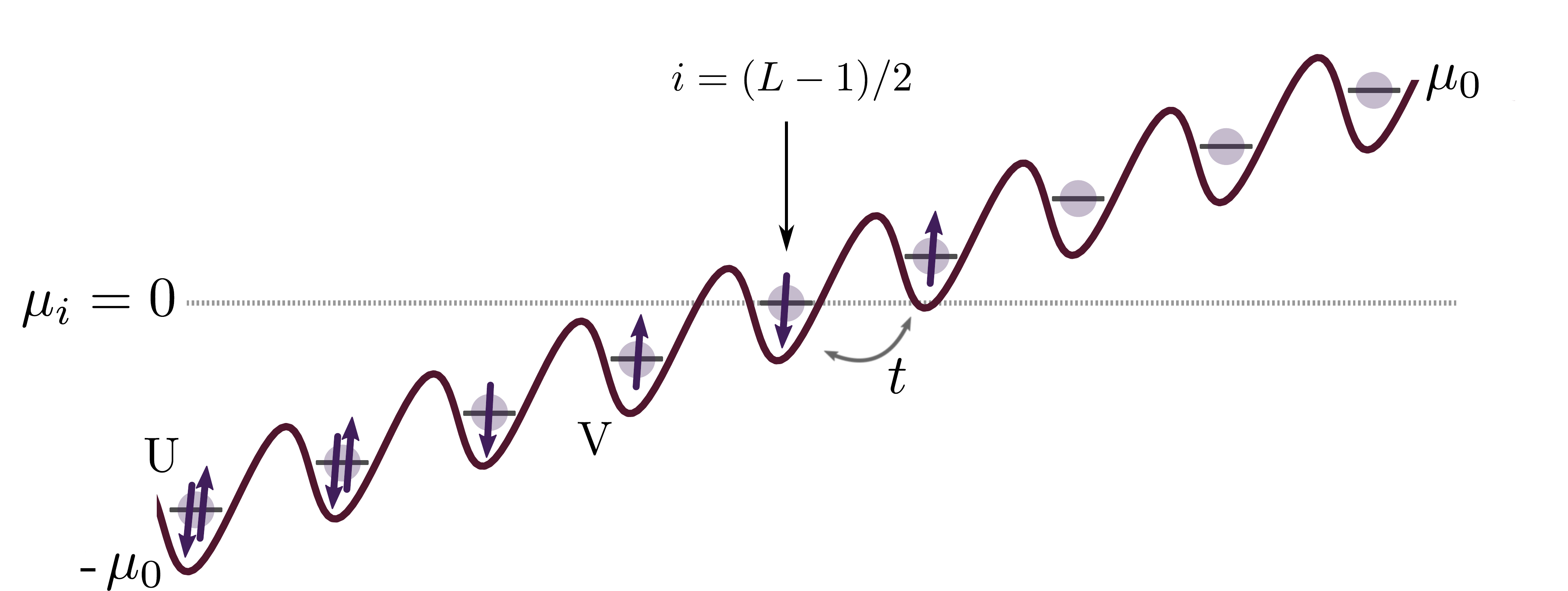}
\caption{Schematic representation of the system considered, for $L=9$ sites.}
\label{system}
\end{figure}

%%%%%%%%%%%%%%%%%%%%%%%%%%%%%%%%%%%%%%%%%%%%%%%%%%%%%
%%%%%%%%%%%%%%%%%%%%%%%%%%%%%%%%%%%%%%%%%%%%%%%%%%%%%

\vspace{0.2cm}

\noindent{\it Results.---} Unless otherwise stated,  we use $U/t\rightarrow U=10$, $\mu=0$ and $L=41$. All energies are measured in units of $t$.
The Hamiltonian is diagonalized exploiting   the
DMRG technique, with around $m=500$ states and 4 to 6 finite-size sweeps, which leads to a precision of around $10^{-10}$ in the energy. For a very steep potential ($\mu^{}_0\rightarrow\infty$) we obtain only two coexisting `phases': a completely filled band ($n^{}_i=2$) up to the center point $i^{}_c$, and completely empty sites ($n^{}_i=0$) above that point, as expected. Both regions are incompressible and insulating. As the slope $\mu^{}_0$ decreases (but remains large), these two `phases' remain near the two ends of the system, but new structures (`phases') appear between them, in which $\langle n^{}_i\rangle$ decreases gradually from $2$ to $0$. Figure \ref{static} presents typical results, for three values of $V$.
Note the electron-hole symmetry between the two sides of Figs. \ref{static}(a-c), which follows directly from Eq. (\ref{hamil}) at $\mu=0$.

%%%%%%%%%%%%%%%%%%%%%%%%%%%%%%%%%%%%%%%%%%%%%%%%%%%%%
%%%%%%%%%%%%%%%%%%%%%%%%%%%%%%%%%%%%%%%%%%%%%%%%%%%%%

\begin{figure}
\includegraphics[width=0.48\textwidth]{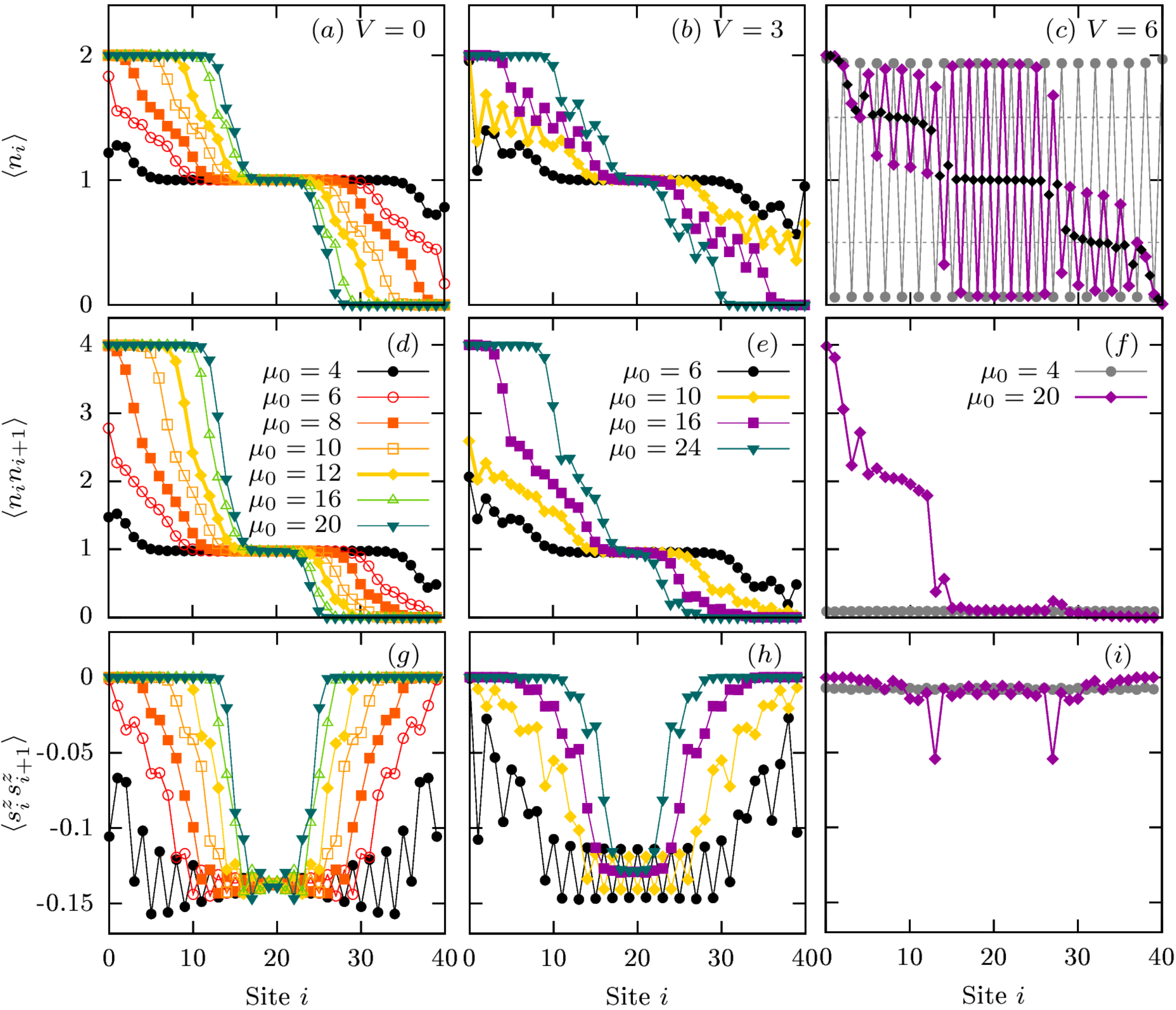}
\caption{(color online) (a)-(c): The local  density $\langle n_i\rangle$; (d)-(f): the nn density-density correlations $\langle n_in_{i+1}\rangle$;  (g)-(i):  the nn spin-spin correlations $\langle s^z_is^z_{i+1}\rangle$, for $V=0,3,6$ and different values of $\mu_0$. The black diamonds in (c) indicate the mean value between neighboring sites. }
\label{static}
\end{figure}

%%%%%%%%%%%%%%%%%%%%%%%%%%%%%%%%%%%%%%%%%%%%%%%%%%%%%
%%%%%%%%%%%%%%%%%%%%%%%%%%%%%%%%%%%%%%%%%%%%%%%%%%%%%

For $V=0$ (i.e., the simplest Hubbard Hamiltonian, left column in Fig.~\ref{static}), the system shows the following phases: for large (but finite) values of $\mu^{}_0$ it is a band insulator at both extremes,  completely filled  on the left and  completely empty  on the right. In the region located symmetrically around the center point $i^{}_c$, we find a Mott-insulating state (one particle per site, $\langle n^{}_i\rangle=1$),  and an antiferromagnetic spin-spin correlation function, Fig. \ref{static}(g). As seen in this figure, the  spin correlation function, $\langle s^z_i s^{z}_{i+1}\rangle\simeq-0.14$ (note: $s_i^z\equiv (n^{}_{i,\uparrow}-n^{}_{i,\downarrow})/2$, the $z-$direction is arbitrarily chosen), agrees with its value of the infinite Mott phase~\cite{whiteprb}.  The three insulating commensurate phases are separated by IIF regions with very small but finite gaps, see Fig.~\ref{compress}. These regions differ from the compressible regions found in Ref.~\cite{HF}, possibly because Ref.~\cite{HF} explores 2D systems using the mean-field approximation.
As $\mu^{}_0$ decreases, the band insulating phases on both ends disappear and the Mott region grows, as estimated below. These results are also consistent with the behavior of the density-density  correlations, which vary between $4$ on the left, via $1$ in the Mott phase, to $0$ on the right, Fig.~\ref{static}(d).

For $V=3$ (middle column in Fig.~\ref{static}) the above three insulating `phases' are supplemented by
two regions with an incipient (doped) CDW order on the two sides of the Mott `phase', with local mean fillings `quasi-plateaus' around $\overline{\langle ^{}n_i\rangle} \simeq 1.5$ and $\overline{\langle n^{}_i\rangle} \simeq 0.5$ (quarter filling of holes and of electrons, respectively). The bar indicates a local average over a few sites.
 Unlike the uniform case $\mu^{}_i=0$, the local average fillings in these regions are not exactly $1.5$ and $0.5$. Rather, they can be fitted by $\langle n^{}_i\rangle =A-Bi+C\cos(i\pi)$ (note that $i$ is the site number!). The oscillating term corresponds to a CDW, with a wave vector $q=\pi$  (our lattice constant is $1$) and  structures $212121\dots$ or $101010\dots$ \cite{Com}. However, the term $-Bi$ represents a linear decrease of the actual average, presumably in response to the linear potential. Without this linear `background', such a CDW is consistent with the results of the density-density and spin-spin correlations and with previous results for the doped (non-half-filled) 1D extended Hubbard model~\cite{nakamura} in a uniform potential, $\mu^{}_i=0$, for which there is a transition from a Tomonaga-Luttinger liquid to a CDW phase for intermediate values of $2t\leq V<U/2$ and large values of $U$ ($U\gg t$). In those cases this CDW phase is insulating and incompressible.
As we discuss below, we also find that, in spite of the varying average local densities, the local density of states has a (small) gap at the Fermi energy, which is consistent with an incompressible state.
 As before, when $\mu^{}_0$ decreases, the Mott region grows, the incipient CDW regions move towards the boundaries and the band-insulating regions disappear.

For $V= 6$ (right column in Fig.~\ref{static})  the Mott region disappears and is replaced by a half-filled CDW, $202020\dots$. For large $\mu^{}_0$'s this phase exists in the center and coexists with doped CDW's at both sides, with fillings $\overline{\langle ^{}n_i\rangle} \simeq 1.5$ and $\overline{\langle n^{}_i\rangle} \simeq 0.5$ respectively (black diamonds in Fig.~\ref{static}(c)). This coexistence of two different CDW's has not been seen before and constitutes a situation which could be observed in cold-atom experiments.  As before, the doped CDW's are accompanied by a very small gradual decrease of the local average occupation -`quasi-plateaus', presumably due to the slope in the potential. When $\mu^{}_0$ is lowered, the half-filled CDW occupies the whole chain. This is expected, since it is well known that when $V>U/2$ and for a half-filled system, the uniform chain undergoes a transition from a Mott phase to a CDW~\cite{Glocke,nakamura}. The results are consistent with the behavior of the density-density and spin-spin correlations. It is interesting to see a finite value of the spin-spin correlations at the phase boundaries between the half-filled and doped CDW's. It is also interesting to see that for $V=3$ the average occupation $\overline{\langle n^{}_i\rangle}$, and the amplitude of the incipient CDW decrease gradually towards the central Mott or CDW region, but this decrease becomes abrupt for $V=6$.
The width of the IIF  region (domain wall) between the two CDW phases seems to shrink to zero above some `critical' value of $V$.

The above results exhibited `plateaus' only for $1/2$, $1/4$ and $3/4$ fillings. We expect similar `plateaus', corresponding to other simple fraction, e.g., $1/8$. However, to see these one would need a much larger number of sites, and this is not possible with our present computer capabilities. Note, though,  that calculations with a smaller number of sites do still show similar steps  for these commensurate fillings.

%%%%%%%%%%%%%%%%%%%%%%%%%%%%%%%%%%%%%%%%%%%%%%%%%%%%%
%%%%%%%%%%%%%%%%%%%%%%%%%%%%%%%%%%%%%%%%%%%%%%%%%%%%%

\begin{figure}[h]
\centering
 \includegraphics[width=0.48\textwidth]{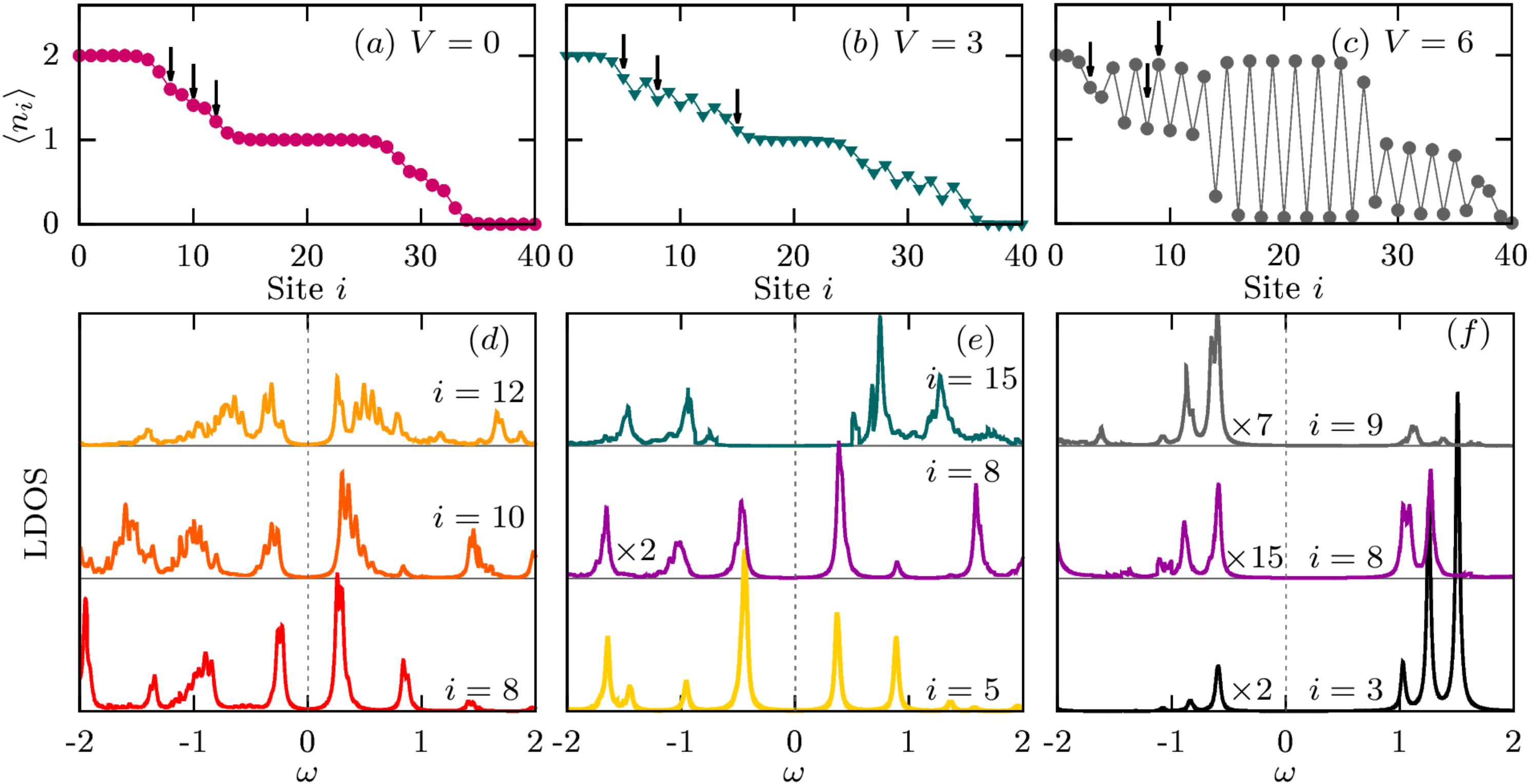}
\caption{(color online) { Top: Local density profile $\langle n_i\rangle$ showing the sites where the local density of states ( LDOS) has been calculated, for $V=0$
($ \mu^{}_0=10$), $V=3$ ($ \mu^{}_0=16$) and $V=6$ ($ \mu^{}_0=20$). Bottom: LDOS showing gaps at the Fermi energy (at $\omega=E^{}_F=0$) for all cases, using $\eta=0.01$ (Eq.~S1 in~\cite{SM}) .
}
}
\label{compress}
\end{figure}

%%%%%%%%%%%%%%%%%%%%%%%%%%%%%%%%%%%%%%%%%%%%%%%%%%%%%
%%%%%%%%%%%%%%%%%%%%%%%%%%%%%%%%%%%%%%%%%%%%%%%%%%%%%

\vspace{0.2cm}

\noindent{\it Local Density of States.---} To further explore  the different phases, we have calculated the local, site dependent, density of states (LDOS), using  the lesser and greater Green's functions, see details in Ref. \cite{SM}. In Fig.~\ref{compress} we show the LDOS for particular sites of the chain for different parameters. We observe that there is always a gap at $E^{}_F=0$, even for the partially filled sites (we have added the filling profile for comparison). The gaps corresponding to these sites are smaller than the corresponding gaps of the fully formed CDW (see the $V=6$ case) and much smaller than those of the Mott region (see Fig.~\ref{dosV0}).  These gaps indicate that these regions are incompressible (non-metallic). This is not a finite size effect (since we would have a finite LDOS at $E^{}_F$ for fractional densities), but a consequence of the linear potential.
We also observe that the LDOS consists of a series of peaks separated by minigaps, a possible indication of Stark discretization~\cite{stark}.

Figure \ref{dosV0} shows a heatplot of the local density of states  along the chain for $V=0$, $\mu=0$ and $\mu^{}_0=10$. The Fermi energy is marked by a white (dashed) line at $\omega=0$.
As the Hamiltonian is particle-hole symmetric around the middle of the chain, the density of states  for the right half of the chain ($20\leq i \leq 40$, not shown) is inverted as a function of $\omega$    (details see in \cite{SM}).
 As mentioned above (Fig.~\ref{compress}), we always find a gap at $E^{}_F$, indicating an incompressible state. This gap is more than an order of magnitude smaller than the Mott gap. We also see a structure in the Hubbard bands in the form of three main substructures which evolve along the chain sites. Each substructure extends to around three neighboring sites, also an indication of Stark localization which requires future study~\cite{stark}.

%%%%%%%%%%%%%%%%%%%%%%%%%%%%%%%%%%%%%%%%%%%%%%%%%%%%%
%%%%%%%%%%%%%%%%%%%%%%%%%%%%%%%%%%%%%%%%%%%%%%%%%%%%%

\begin{figure}[h!]
\centering
\includegraphics[width=0.48\textwidth]{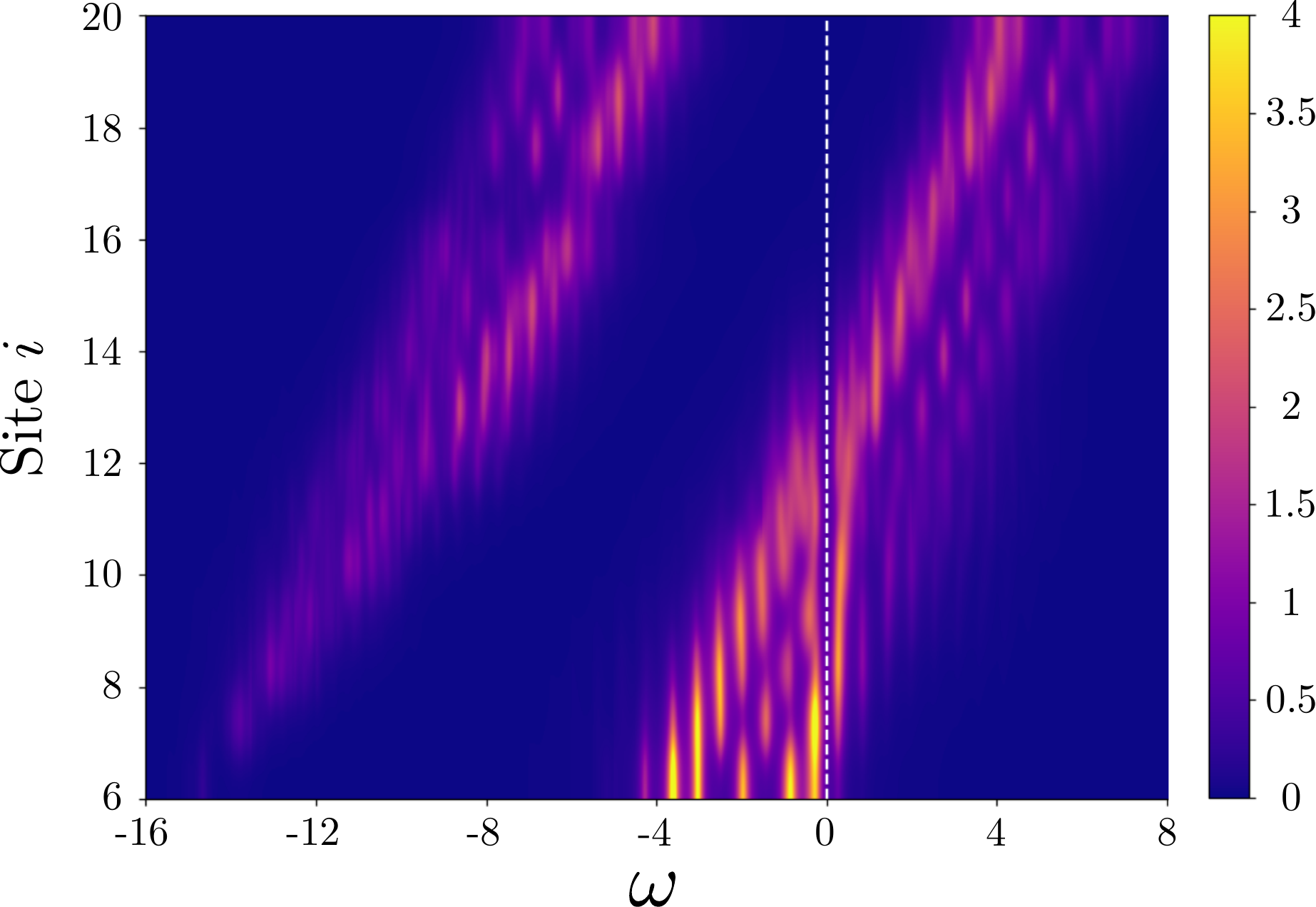}
\caption{Heatplot of the local density of states at different sites ($6\leq i \leq 20$) for $\mu^{}_0=10$ and $V=0$. The Fermi energy is marked by a white (dashed) line at $\omega=0$}
\label{dosV0}
\end{figure}

%%%%%%%%%%%%%%%%%%%%%%%%%%%%%%%%%%%%%%%%%%%%%%%%%%%%%
%%%%%%%%%%%%%%%%%%%%%%%%%%%%%%%%%%%%%%%%%%%%%%%%%%%%%

An interesting result for $V=0$ is the existence of a (negative) high-energy localized state in the IIF region (clearly seen in the density of states plots at the left of the chain, Fig. S2 in Ref. \cite{SM}). We can see a small and narrow peak at energies around $\omega \sim-14$ for the first sites of this region, which evolves to higher energies (following the increase of $\mu$),  while we approach the Mott region, increasing its width. This state is reminiscent of the lower Hubbard band for the left  regions. A similar state is seen for the right half of the chain which is reminiscent of the upper Hubbard band (not shown).  More results for the density of states, together with some calculations in the atomic limit, are presented in Ref. \cite{SM}.

%%%%%%%%%%%%%%%%%%%%%%%%%%%%%%%%%%%%%%%%%%%%%%%%%%%%%
%%%%%%%%%%%%%%%%%%%%%%%%%%%%%%%%%%%%%%%%%%%%%%%%%%%%%

\vspace{0.2cm}

\noindent{\it Size of the Mott region.---} At the electron-hole symmetric case (and $V=0$) the upper and lower Hubbard bands are centered at $\pm \frac{U}{2}$ respectively, each with a total width of 4.
For $\mu=0$, the size of the Mott region can be estimated recalling that the Mott insulating state requires that the local $\mu_{i}$ lies within the Mott gap i.e., $-U/2+2 < \mu^{}_i < U/2-2$. At the lower limit $\mu^{}_{\rm min}=-\frac{U}{2}+2$, yielding by Eq. (\ref{mui})
that $i^{}_{\rm min}\mu^{}_{0}=i^{}_c(\mu^{}_{0}-\frac{U}{2}+2)$,
while at $\mu^{}_{\rm max}=\frac{U}{2}-2$ one finds
$i^{}_{\rm max}\mu^{}_{0}=i^{}_c(\mu^{}_{0}+\frac{U}{2}-2)$. Consequently, assuming that the width of the Hubbard bands is not modified by the presence of the confining potential, the size of the Mott region is:
\begin{align}
L^{}_{\rm Mott}=i^{}_{\rm max}-i^{}_{\rm min}=\left( U/2 - 2\right)(L-1)/\mu^{}_0\ .
\label{Lmott}
\end{align}
As the confining potential slightly increases the width of the Hubbard bands (not shown), the gap in-between them and $L^{}_{\rm Mott}$ are slightly overestimated.

To compare Eq. (\ref{Lmott}) with our numerical results, we have estimated the size of the Mott region by
defining its boundaries at the points where the linear fits of the numerical derivative of the local occupation
intercept $0$ for each value of  $\mu^{}_0$, using the results shown in Fig. \ref{static}(a). This procedure reveals that indeed  the size of the Mott region is proportional to $1/\mu^{}_0$ (see Fig. S1 in Ref. \cite{SM}), and it shrinks to zero for very steep potentials.

%%%%%%%%%%%%%%%%%%%%%%%%%%%%%%%%%%%%%%%%%%%%%%%%%%%%%
%%%%%%%%%%%%%%%%%%%%%%%%%%%%%%%%%%%%%%%%%%%%%%%%%%%%%

\vspace{0.2cm}

\noindent{\it Changing the global chemical potential.---} The coexisting phases are robust against changes in the global chemical potential $\mu$. In Fig. \ref{HBmodif} we show our results for two cases, $V=0$ (with coexisting band and Mott insulators, separated by intermediate IIF regions), and $V=4$ (with CDW's and Mott insulators).
The different phases shift towards the right or the left with respect to their position for $\mu=0$ but otherwise are not changed, except for the regions close to the boundaries where they are affected by the open boundaries.

%%%%%%%%%%%%%%%%%%%%%%%%%%%%%%%%%%%%%%%%%%%%%%%%%%%%%
%%%%%%%%%%%%%%%%%%%%%%%%%%%%%%%%%%%%%%%%%%%%%%%%%%%%%

\begin{figure}[h]
\centering
\includegraphics[width=0.45\textwidth]{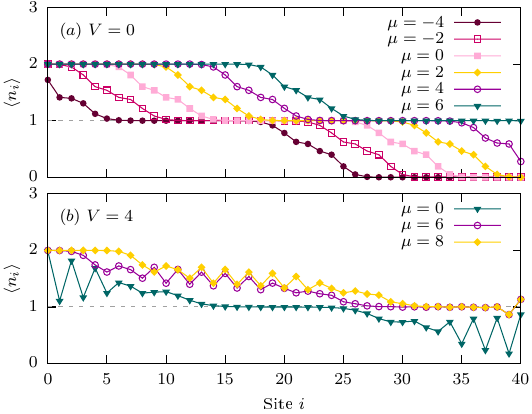}
\caption{(color online) The local density at $\mu^{}_0=10$, for $V=0$ (a), and $V=4$ (b), for different values of the global chemical potential $\mu$.}
\label{HBmodif}
\end{figure}

%%%%%%%%%%%%%%%%%%%%%%%%%%%%%%%%%%%%%%%%%%%%%%%%%%%%%
%%%%%%%%%%%%%%%%%%%%%%%%%%%%%%%%%%%%%%%%%%%%%%%%%%%%%

\vspace{0.2cm}

\noindent{\it Discussion.---} In this paper we study the one-dimensional extended Hubbard model, subject to a linearly-varying Wannier-Stark  potential on a finite chain,  applying the density-matrix renormalization group. We find an interesting sequence of several insulating electronic phases in the ground state, in which regions with commensurate charge density waves coexist with band and Mott insulating phases. These regions are separated by incompressible domain walls with incommensurate fillings, which were not reported before. The results are summarized in Fig. ~\ref{PD}.
 Further research is needed to define whether these incompressible walls are due to the Stark many-body localization~\cite{moessner}.
The steeper the slope of the external potential, the narrower the domain walls. These phases and domain walls can be moved around by varying a global chemical potential, thus providing a possible functionality of this kind of systems. Cold-atom chains placed in an external electric field are suggested as experimental realizations of our system.

%%%%%%%%%%%%%%%%%%%%%%%%%%%%%%%%%%%%%%%%%%%%%%%%%%%%%
%%%%%%%%%%%%%%%%%%%%%%%%%%%%%%%%%%%%%%%%%%%%%%%%%%%%%

\begin{figure}[h!]
\centering
\includegraphics[width=0.5\textwidth]{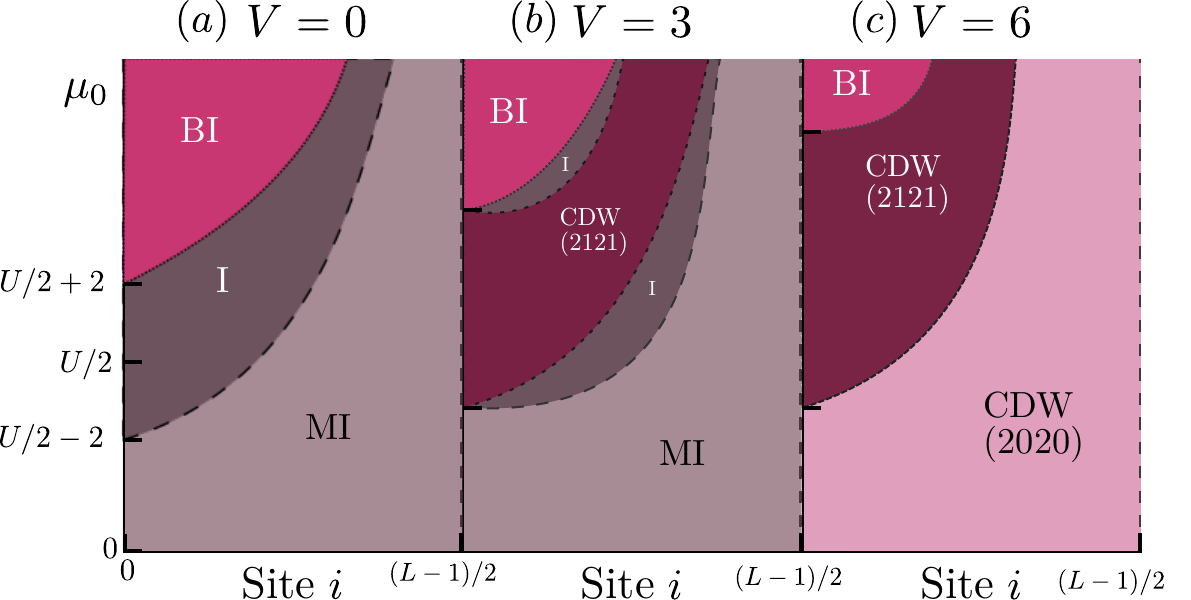}
\caption{Schematic plot of the various phases observed in the chain subjected to a linear potential,  for three different regimes defined by the nearest-neighbor interaction,  $V$. The sequence of numbers, 212121 and 202020, refer to the occupations of neighboring sites. MI: Mott insulator, BI: band insulator, IIF: incompressible incommensurate-filling phase. }
\label{PD}
\end{figure}

%%%%%%%%%%%%%%%%%%%%%%%%%%%%%%%%%%%%%%%%%%%%%%%%%%%%%
%%%%%%%%%%%%%%%%%%%%%%%%%%%%%%%%%%%%%%%%%%%%%%%%%%%%%

\vspace{0.3cm}

\noindent{\it Acknowledgments.---} NAB and KH acknowledge support from ICTP through the STEP and Associates Programmes respectively, and from the PICT 2018-01546 grant of the ANPCyT. The authors thank useful discussions with Carlos Balseiro.

%%%%%%%%%%%%%%%%%%%%%%%%%%%%%%%%%%%%%%%%%%%%%%%%%%%%%%

\newpage

\newpage

\widetext
\begin{center}

%-------------------------------------------------------------------------------------------------------------------------------------------------------------------------------

\section{Supplemental material for "Coexistence of insulating phases in confined \\fermionic chains with a Wannier-Stark potential"}
\end{center}
\setcounter{equation}{0}
\setcounter{figure}{0}
\setcounter{table}{0}
\setcounter{page}{1}
\makeatletter
\renewcommand{\theequation}{S\arabic{equation}}
\renewcommand{\thefigure}{S\arabic{figure}}
\renewcommand{\bibnumfmt}[1]{[S#1]}
\renewcommand{\citenumfont}[1]{S#1}

%%%%%%%%%%%%%%%%%%%%%
%%%%%%%%%%%%%%%%%%%%%%%%%%%%%%%%%%%%%%%%%%%%%%%%%%%%%
\author{N. Aucar Boidi}
\affiliation{Centro At\'{o}mico Bariloche, Instituto Balseiro, 8400 Bariloche, Argentina}
\email{nairaucar@gmail.com}

\author{K. Hallberg}
\affiliation{Centro At\'{o}mico Bariloche, Instituto Balseiro, 8400 Bariloche, Argentina}
\affiliation{Instituto de Nanociencia y Nanotecnolog\'{\i}a CNEA-CONICET, 8400 Bariloche, Argentina}
\email{karenhallberg@gmail.com}
%%%%%%%%%%%%%%%%%%%%%%%%%%%%%%%%%%%%%%%%%%%%%%%%%%%%%

\author{Amnon Aharony}
\affiliation{School of Physics and Astronomy, Tel Aviv University, Tel Aviv 6997801, Israel}
\email{aaharonyaa@gmail.com}

\author{Ora Entin-Wohlman}
\affiliation{School of Physics and Astronomy, Tel Aviv University, Tel Aviv 6997801, Israel}
\email{orawohlman@gmail.com}

\date{\today}

\maketitle

%%%%%%%%%%%%%%%%%%%%%%%%%%%%%%%%%%%%%%%%%%%%%%%%%%%%%%%
%%%%%%%%%%%%%%%%%%%%%%%%%%%%%%%%%%%%%%%%%%%%%%%%%%%%%%
\noindent{\it Size of the Mott and band insulator regions.---} Figure \ref{Lmott} presents  the  size of the Mott region as a function of $1/\mu^{}_{0}$ for two different values of $U$ and $L$ and for $V=0$, as a function of $1/\mu^{}_0$. It confirms the linear dependence on $1/\mu^{}_{0}$, as found in Eq. (3) of the main text.

%%%%%%%%%%%%%%%%%%%%%%%%%%%%%%%%%%%%%%%%%%%%%%%%%%%%%%%
%%%%%%%%%%%%%%%%%%%%%%%%%%%%%%%%%%%%%%%%%%%%%%%%%%%%%%

\begin{figure}[h]
\centering
 \includegraphics[width=0.48\textwidth]{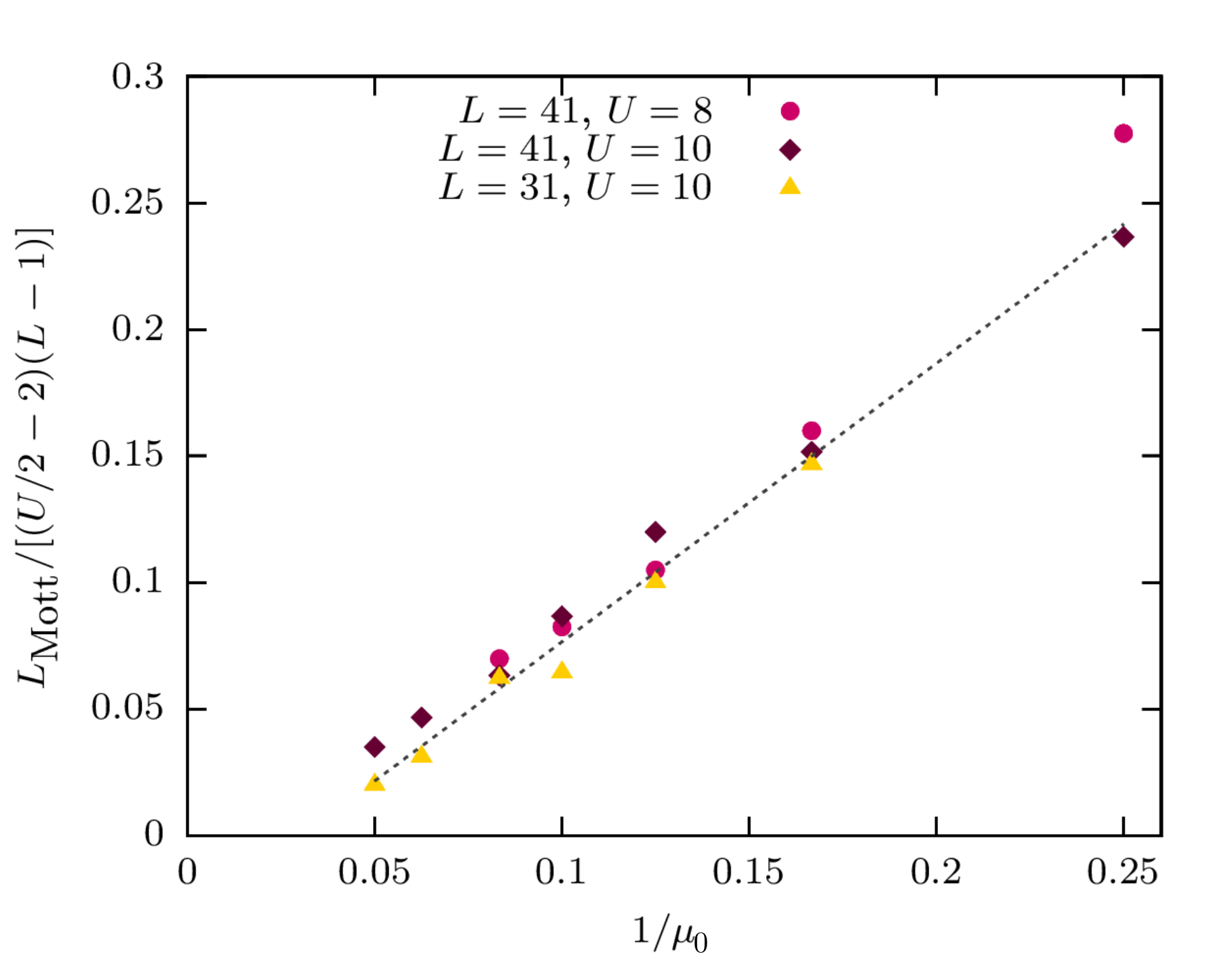}
 \caption{The size of the Mott region, scaled by $(U/2-2)(L-1)$ for  various values of  $L$ and $V$. The dotted  line is a guide to the eye, indicating the linear dependence on $1/\mu^{}_{0}$.}
 \label{Lmott}
\end{figure}

%%%%%%%%%%%%%%%%%%%%%%%%%%%%%%%%%%%%%%%%%%%%%%%%%%%%%%%
%%%%%%%%%%%%%%%%%%%%%%%%%%%%%%%%%%%%%%%%%%%%%%%%%%%%%%

\noindent{\it Local density of states.---}
Introducing  the lesser and greater Green's functions at site $j$,
\begin{align}
A^{>}_{j\sigma}(\omega)&=-\frac{1}{\pi}\text{Im}\langle c^{}_{j\sigma}(\omega+i\eta-H+E^{}_{0})^{-1}c^{\dagger}_{j\sigma}\rangle\ ,\nonumber\\
A^{<}_{j\sigma}(\omega)&=-\frac{1}{\pi}\text{Im}\langle c_{j\sigma}^{\dagger}(\omega+i\eta-H+E^{}_{0})^{-1}c^{}_{j\sigma}\rangle\ ,\label{Aq}
\end{align}
where the expectation value refers to the ground state of the system whose energy is $E^{}_0$  and $\eta$ is a small Lorentzian  broadening, the local density of states is given by
\begin{align}
A^{}_{j}\left(\omega\right) = \sum_{\sigma} \big[ A^{>}_{j\sigma}\left(\omega\right) + A^{<}_{j\sigma}\left(-\omega\right)\big]\ .
\label{DOS}
\end{align}
This quantity is found numerically, exploiting the density-matrix renormalization group technique.

Below we show the detailed local density of states calculated for $U=10$, $V=0$ and $V=8$ (all energies are measured in units of $t$ and relative to the Fermi energy $E_F$, see the main text).
\begin{enumerate}
\item
$V=0$: Figure \ref{V0} shows the local density of states at different sites. The upper (UHB) and lower (LHB) Hubbard bands are clearly seen, separated by an energy $~U$. The UHB  crosses the Fermi energy for sites $6\leq i \leq 14$, but it always displays a small gap at $\omega=0$,
(more details for a smaller $\eta=0.01$ are given  in Fig. 3 of the main text), indicating
an incompressible behavior in the presence of the Wannier-Stark potential, unlike the metallic behavior observed in the uniform case
(Fig. 3 in the main text). For $14< i \leq 20$ the Fermi energy lies within the Mott gap and this sector of the chain is a Mott insulator (Fig. 2 in the main text). It is particle-hole symmetric for $20\leq i \leq 40$.
The wiggles are due to the finite size of each region. It is interesting to note that the density of states at $\omega < E_F=0$ at $i=6$ resembles the free particle situation. This is due to the fact that the system is nearly completely filled at that site and the dynamics of one hole can be understood by that of a nearly free particle. Also interesting is the evolution of the LHB which begins with a very low weight at low energies (close to $\omega=-15$ for $i=6$) and acquires more weight and gets wider towards the middle of the chain.

%%%%%%%%%%%%%%%%%%%%%%%%%%%%%%%%%%%%%%%%%%%%%%%%%%%%%%%
%%%%%%%%%%%%%%%%%%%%%%%%%%%%%%%%%%%%%%%%%%%%%%%%%%%%%%

\begin{figure}[h]
\centering
\includegraphics[width=0.5\textwidth]{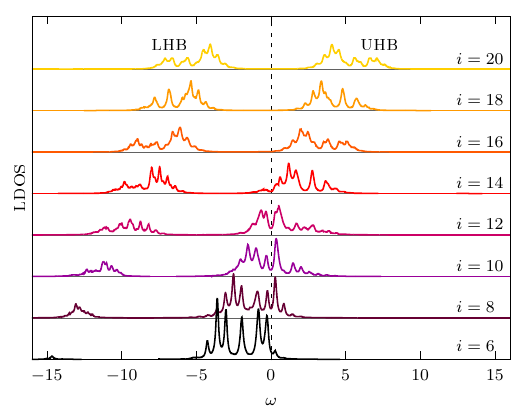}
\caption{Local density of states at different sites (shifted vertically) for $\mu^{}_0=10$, $V=0$, $\eta=0.1$ and $L=41$ sites. The Fermi energy is $E^{}_F=0$}
\label{V0}
\end{figure}

%%%%%%%%%%%%%%%%%%%%%%%%%%%%%%%%%%%%%%%%%%%%%%%%%%%%%%%
%%%%%%%%%%%%%%%%%%%%%%%%%%%%%%%%%%%%%%%%%%%%%%%%%%%%%%

\begin{figure}[h!]
\centering
\includegraphics[width=0.5\textwidth]{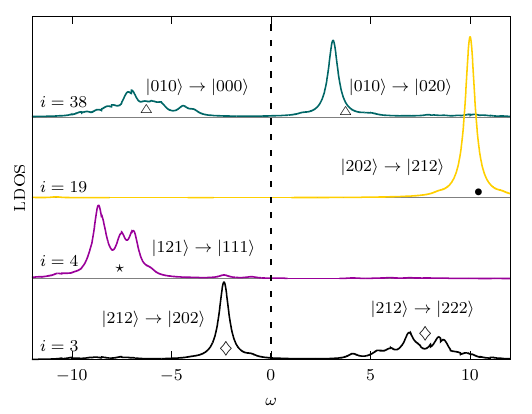}
 \caption{Local density of states for $\mu^{}_0=16$, $V=8$ and $L=41$ sites at different sites of the chain (shifted vertically). Symbols indicate excitations energies for different processes. }
 \label{V4}
\end{figure}

%%%%%%%%%%%%%%%%%%%%%%%%%%%%%%%%%%%%%%%%%%%%%%%%%%%%%%%
%%%%%%%%%%%%%%%%%%%%%%%%%%%%%%%%%%%%%%%%%%%%%%%%%%%%%%

\item

Case $V=8$: Figure \ref{V4} presents  the density of states for the different CDW regions, all showing a gap at the Fermi energy.
To understand these results we calculate the energy of having different configurations in one atom taking into account its nearest neighbors ($t=0$) and adding/removing one particle at the atomic (middle) site $i$. The different states are encoded in the three sites as $(n_1, n_2, n_3 )$, where $n_i=0,1,2$ indicates the local charge. The corresponding transitions are marked in Fig. \ref{V4} and their energies are calculated below. We use $\mu^{}_0=16$, $V=8$ and $U=10$.

\end{enumerate}

\begin{itemize}
\item \textit{$i=3$ (diamonds in Fig. \ref{V4}):}

\begin{eqnarray}
E(222)-E(212)&=&U/2-\mu+\mu^{}_i+2V=7.4 \nonumber \\
E(202)-E(212)&=&U/2+\mu -\mu_i-2V=-2.6 \nonumber \\ \nonumber
\end{eqnarray}

\item \textit{$i=4$ (star):}

\begin{equation}
E(121)-E(111)=U/2-\mu+\mu_i=-7.8 \nonumber \\
\end{equation}

\item \textit{$i=19$ (black dot):}

\begin{equation}
E(212)-E(202)=-U/2-\mu +\mu_i+2V=10.2 \nonumber \\
\end{equation}

\item \textit{$i=38$ (empty triangles):}

\begin{eqnarray}
E(020)-E(010)&=&U/2-2V-\mu+\mu_i=3.4 \nonumber \\
E(010)-E(000)&=&-U/2-2V-\mu+\mu_i=-6.6 \nonumber \\ \nonumber
\end{eqnarray}

\end{itemize}

%-------------------------------------------------------------------------------------------------------------------------------------

\begin{thebibliography}{99}
%%%%%%%%%%%%%%%%%%%%%%%%%%%%%%%%%%%%%%%%%%%%%%%%%%%%%
\bibitem{hubbard}
F. H. L. Essler, H. Frahm, F. G\"ohmann, A. Kl\"umper, and V. E. Korepin,
{\it The one-dimensional Hubbard model}, \href{https://doi.org/10.1017/CBO9780511534843}{(Cambridge University Press, 2009)}.
%%%%%%%%%%%%%%%%%%%%%%%%%%%%%%%%%%%%%%%%%%%%%%%%%%%%%%
%%%%%%%%%%%%%%%%%%%%%%%%%%%%%%%%%%%%%%%%%%%%%%%%%%%%%
\bibitem{Anderson}
P. W. Anderson,
{\it Antiferromagnetism: Theory of Superexchange Interaction,}
\href{https://journals.aps.org/pr/abstract/10.1103/PhysRev.79.350}{Phys. Rev. {\bf 79}, 350 (1950)}.
%%%%%%%%%%%%%%%%%%%%%%%%%%%%%%%%%%%%%%%%%%%%%%%%%%%%%%
%%%%%%%%%%%%%%%%%%%%%%%%%%%%%%%%%%%%%%%%%%%%%%%%%%%%%
\bibitem{Mott}
T. Giamarchi,
{\it Mott transition in one dimension},
\href{https://doi.org/10.1016/S0921-4526(96)00768-5}{Physica B: Condensed Matter {\bf 230-232},  975 (1997)}.
%%%%%%%%%%%%%%%%%%%%%%%%%%%%%%%%%%%%%%%%%%%%%%%%%%%%%%
%%%%%%%%%%%%%%%%%%%%%%%%%%%%%%%%%%%%%%%%%%%%%%%%%%%%%
\bibitem{Mila}
F. Mila and X. Zotos,
{\it Phase Diagram of the One-Dimensional Extended Hubbard Model at Quarter-Filling},
\href{https://iopscience.iop.org/article/10.1209/0295-5075/24/2/010}{Euro. Phys. Lett. {\bf 24}, 133 (1993)}.
%%%%%%%%%%%%%%%%%%%%%%%%%%%%%%%%%%%%%%%%%%%%%%%%%%%%%%
%%%%%%%%%%%%%%%%%%%%%%%%%%%%%%%%%%%%%%%%%%%%%%%%%%%%%
\bibitem{Zhang}
Y. Z. Zhang,
{\it Dimerization in a Half-Filled One-Dimensional Extended Hubbard Model},
\href{https://journals.aps.org/prl/abstract/10.1103/PhysRevLett.92.246404}{Phys. Rev. Lett. {\bf 92}, 246404 (2004)}.
%%%%%%%%%%%%%%%%%%%%%%%%%%%%%%%%%%%%%%%%%%%%%%%%%%%%%%
%%%%%%%%%%%%%%%%%%%%%%%%%%%%%%%%%%%%%%%%%%%%%%%%%%%%%
\bibitem{Ejima}
S. Ejima, and S. Nishimoto,
{\it Phase Diagram of the One-Dimensional Half-Filled Extended Hubbard Model},
\href{https://journals.aps.org/prl/abstract/10.1103/PhysRevLett.99.216403}{Phys. Rev. Lett. {\bf 99}, 216403 (2007)}.
%%%%%%%%%%%%%%%%%%%%%%%%%%%%%%%%%%%%%%%%%%%%%%%%%%%%%%
%%%%%%%%%%%%%%%%%%%%%%%%%%%%%%%%%%%%%%%%%%%%%%%%%%%%%
\bibitem{Glocke}
S. Glocke, A. Kl\"{u}mper, and J. Sirker,
{\it Half-filled one-dimensional extended Hubbard model: Phase diagram and thermodynamics,}
\href{https://journals.aps.org/prb/abstract/10.1103/PhysRevB.76.155121}{Phys. Rev. B {\bf 76}, 155121 (2007)}.
%%%%%%%%%%%%%%%%%%%%%%%%%%%%%%%%%%%%%%%%%%%%%%%%%%%%%%
%%%%%%%%%%%%%%%%%%%%%%%%%%%%%%%%%%%%%%%%%%%%%%%%%%%%%

%\bibitem{EHM} D. Baeriswyl, {\it Theoretical Aspects of Conducting Polymers: Electronic Structure and Defect States}, in
%\href{https://link.springer.com/chapter/10.1007/978-94-009-5299-7_1}{{\bf Theoretical Aspects of Band Structures
%and Electronic Properties of Pseudo-One-Dimensional
%Solids}, edited by R. H. Kamimura (Reidel, Dordrecht,
%1985), pp. 1-48.}
%\bibitem{Ishiguro}
%T. Ishiguro,  K. Yamaji, and G. Saito,
%{\it Organic Superconductors}
%(Springer-Verlag, Berlin, 1990).
%\bibitem{2} H. Kiess, ed., {\bf Conjugated Conducting Polymers},
%(Springer-Verlag, Berlin, 1992).
%\bibitem{3} H. Ishii {\it et al.}, {\it Direct observation of Tomonaga-Luttinger-liquid state in carbon nanotubes at low temperatures}, \href{https://www.nature.com/articles/nature02074}{Nature (London) 426, 540 (2003)}.

%%%%%%%%%%%%%%%%%%%%%%%%%%%%%%%%%%%%%%%%%%%%%%%%%%%%%%
%%%%%%%%%%%%%%%%%%%%%%%%%%%%%%%%%%%%%%%%%%%%%%%%%%%%%
\bibitem{bloch}
I. Bloch, J. Dalibard, and W. Zwerger,
{\it Many-body physics with ultracold gases},
\href{https://journals.aps.org/rmp/abstract/10.1103/RevModPhys.80.885}{Rev. Mod. Phys. {\bf 80}, 885 (2008)}. %, Sec. V.
%%%%%%%%%%%%%%%%%%%%%%%%%%%%%%%%%%%%%%%%%%%%%%%%%%%%%%
%%%%%%%%%%%%%%%%%%%%%%%%%%%%%%%%%%%%%%%%%%%%%%%%%%%%%
\bibitem{guan}
X. W. Guan, M. Batchelor, and C. Lee,
{\it Fermi gases in one dimension: From Bethe Ansatz to experiments},
\href{https://journals.aps.org/rmp/abstract/10.1103/RevModPhys.85.1633}{Rev. Mod. Phys. {\bf 85}, 1633 (2013)}.
%%%%%%%%%%%%%%%%%%%%%%%%%%%%%%%%%%%%%%%%%%%%%%%%%%%%%%
%%%%%%%%%%%%%%%%%%%%%%%%%%%%%%%%%%%%%%%%%%%%%%%%%%%%%
\bibitem{HF} U. Khanna, Y. Gefen, O. Entin-Wohlman, and A. Aharony,
{\it Edge Reconstruction of a Time-Reversal Invariant Insulator: Compressible-Incompressible Stripes},
\href{https://journals.aps.org/prl/abstract/10.1103/PhysRevLett.128.186801}
{Phys. Rev. Lett. {\bf 128}, 186801 (2022)} and references therein.
%%%%%%%%%%%%%%%%%%%%%%%%%%%%%%%%%%%%%%%%%%%%%%%%%%%%%%
%%%%%%%%%%%%%%%%%%%%%%%%%%%%%%%%%%%%%%%%%%%%%%%%%%%%%
\bibitem{hirsch}
J. E. Hirsch,
{\it Charge-Density-Wave to Spin-Density-Wave Transition in the Extended Hubbard Model},
\href{https://journals.aps.org/prl/abstract/10.1103/PhysRevLett.53.2327}
{Phys. Rev. Lett. {\bf 53}, 2327 (1984)}.
%%%%%%%%%%%%%%%%%%%%%%%%%%%%%%%%%%%%%%%%%%%%%%%%%%%%%%
%%%%%%%%%%%%%%%%%%%%%%%%%%%%%%%%%%%%%%%%%%%%%%%%%%%%%
\bibitem{Nishino}
T. Nishino,
{\it Charge Excitation Gap of the Extended Hubbard Model,}
\href{https://doi.org/10.1143/JPSJ.61.3651}{J. Phys. Soc. Jpn {\bf 61}, 3651 (1992)}.
%%%%%%%%%%%%%%%%%%%%%%%%%%%%%%%%%%%%%%%%%%%%%%%%%%%%%%
%%%%%%%%%%%%%%%%%%%%%%%%%%%%%%%%%%%%%%%%%%%%%%%%%%%%%
\bibitem{Furusaki}
M. Tsuchiizu and A. Furusaki,
{\it Ground-state phase diagram of the one-dimensional half-filled extended Hubbard model},
\href{https://journals.aps.org/prb/abstract/10.1103/PhysRevB.69.035103}{Phys. Rev. B {\bf 69}, 035103 (2004)}.
%%%%%%%%%%%%%%%%%%%%%%%%%%%%%%%%%%%%%%%%%%%%%%%%%%%%%%
%%%%%%%%%%%%%%%%%%%%%%%%%%%%%%%%%%%%%%%%%%%%%%%%%%%%%
\bibitem{Spalding}
J. Spalding, S-W. Tsai, and D. K. Campbell,
{\it  Critical entanglement for the half-filled extended Hubbard model},
\href{https://journals.aps.org/prb/abstract/10.1103/PhysRevB.99.195445}{  Phys. Rev. B {\bf 99}, 195445 (2019)}.
%%%%%%%%%%%%%%%%%%%%%%%%%%%%%%%%%%%%%%%%%%%%%%%%%%%%%%
%%%%%%%%%%%%%%%%%%%%%%%%%%%%%%%%%%%%%%%%%%%%%%%%%%%%%
\bibitem{phased}
e.g., Y-Y. Xiang, X-J. Liu, Y-H. Yuan, J. Cao, and C-M. Tang,
{\it Doping dependence of the phase diagram in one-dimensional extended Hubbard model: a functional renormalization group study}, \href{https://iopscience.iop.org/article/10.1088/1361-648X/aafd4d/meta}{J. Phys.: Condens. Matter 31 125601 (2019)} and references therein.
%%%%%%%%%%%%%%%%%%%%%%%%%%%%%%%%%%%%%%%%%%%%%%%%%%%%%%
%%%%%%%%%%%%%%%%%%%%%%%%%%%%%%%%%%%%%%%%%%%%%%%%%%%%%
\bibitem{Philoxene}
L. Philoxene, V. H. Dao, and R. Fr\'esard,
{\it Spin and charge modulations of a half-filled extended Hubbard model},
\href{https://journals.aps.org/prb/abstract/10.1103/PhysRevB.106.235131}{Phys. Rev. B {\bf 106}, 235131 (2022)}.
%%%%%%%%%%%%%%%%%%%%%%%%%%%%%%%%%%%%%%%%%%%%%%%%%%%%%%
%%%%%%%%%%%%%%%%%%%%%%%%%%%%%%%%%%%%%%%%%%%%%%%%%%%%%


%\bibitem{step}
%A. Cichy, K. J.  Kapcia, and A. Ptok,
%{\it Phase separations induced by a trapping potential in one-dimensional fermionic systems as a source of core-shell structures}, \href{https://doi.org/10.1038/s41598-019-42044-w}{ Sci. Rep. {\bf 9}, 6719 (2019)}.



%%%%%%%%%%%%%%%%%%%%%%%%%%%%%%%%%%%%%%%%%%%%%%%%%%%%%%
%%%%%%%%%%%%%%%%%%%%%%%%%%%%%%%%%%%%%%%%%%%%%%%%%%%%%
\bibitem{parabola}
A. Recati, P. O. Fedichev, W. Zwerger, and P. Zoller,
{\it Spin-Charge Separation in Ultracold Quantum Gases},
\href{https://journals.aps.org/prl/abstract/10.1103/PhysRevLett.90.020401}{Phys. Rev. Lett. {\bf 90}, 020401 (2003)}.
%%%%%%%%%%%%%%%%%%%%%%%%%%%%%%%%%%%%%%%%%%%%%%%%%%%%%%
%%%%%%%%%%%%%%%%%%%%%%%%%%%%%%%%%%%%%%%%%%%%%%%%%%%%%
\bibitem{heidrich} F. Heidrich-Meisner, G. Orso, and A. E. Feiguin, {\it Phase separation of trapped spin-imbalanced Fermi gases in one-dimensional optical lattices}, \href{https://doi.org/10.1103/PhysRevA.81.053602}{Phys. Rev. A {\bf 81}, 053602, (2010).}
%%%%%%%%%%%%%%%%%%%%%%%%%%%%%%%%%%%%%%%%%%%%%%%%%%%%%%
%%%%%%%%%%%%%%%%%%%%%%%%%%%%%%%%%%%%%%%%%%%%%%%%%%%%%
\bibitem{gross}
M. Boll, T.  A. Hilker, G. Salomon, A. Omran, J. Nespolo, L. Pollet, I. Bloch, and C. Gross,
{\it Spin- and density-resolved microscopy of antiferromagnetic correlations in Fermi-Hubbard chains},
\href{https://www.science.org/doi/10.1126/science.aag1635}{Science {\bf 353}, 1257 (2016)}.
%%%%%%%%%%%%%%%%%%%%%%%%%%%%%%%%%%%%%%%%%%%%%%%%%%%%%%
%%%%%%%%%%%%%%%%%%%%%%%%%%%%%%%%%%%%%%%%%%%%%%%%%%%%%
\bibitem{greiner}
D. Greif, M. F. Parsons, A. Mazurenko, C. S. Chiu, S. Blatt, F. Huber, G. Ji, and M. Greiner,
{\it Site-resolved imaging of a fermionic Mott insulator},
\href{https://www.science.org/doi/10.1126/science.aad9041}{Science {\bf 351}, 953 (2016)}.
%%%%%%%%%%%%%%%%%%%%%%%%%%%%%%%%%%%%%%%%%%%%%%%%%%%%%%
\bibitem{stark} M. Udono, T. Kaneko, and K. Sugimoto, {\it Wannier-Stark Ladders and Stark Shifts of Excitons in Mott Insulators}, \href{https://doi.org/10.1103/PhysRevB.108.L081304}
{Phys. Rev. B {\bf 108}, L081304, (2023).}
%%%%%%%%%%%%%%%%%%%%%%%%%%%%%%%%%%%%%%%%%%%%%%%%%%%%%

\bibitem{white}
S. R. White,
{\it Density matrix formulation for quantum renormalization groups},
\href{https://journals.aps.org/prl/abstract/10.1103/PhysRevLett.69.2863}{Phys. Rev. Lett. {\bf 69}, 2863 (1992)}.
%%%%%%%%%%%%%%%%%%%%%%%%%%%%%%%%%%%%%%%%%%%%%%%%%%%%%%
%%%%%%%%%%%%%%%%%%%%%%%%%%%%%%%%%%%%%%%%%%%%%%%%%%%%%
\bibitem{whiteprb}
S. R. White,
{\it Density-matrix algorithms for quantum renormalization groups},
\href{https://journals.aps.org/prb/abstract/10.1103/PhysRevB.48.10345}{Phys. Rev. B {\bf 48}, 10345 (1993)}.
%%%%%%%%%%%%%%%%%%%%%%%%%%%%%%%%%%%%%%%%%%%%%%%%%%%%%%
%%%%%%%%%%%%%%%%%%%%%%%%%%%%%%%%%%%%%%%%%%%%%%%%%%%%%
\bibitem{reviewkaren}
 K. A. Hallberg,
 {\it New trends in density matrix renormalization},
 \href{https://www.tandfonline.com/doi/full/10.1080/00018730600766432}{ Adv. Phys. {\bf 55}, 477 (2006)}.
 %%%%%%%%%%%%%%%%%%%%%%%%%%%%%%%%%%%%%%%%%%%%%%%%%%%%%%
%%%%%%%%%%%%%%%%%%%%%%%%%%%%%%%%%%%%%%%%%%%%%%%%%%%%%
\bibitem{Uli} U. Schollw\"ock,
{\it The density-matrix renormalization group}, \href{https://journals.aps.org/rmp/abstract/10.1103/RevModPhys.77.259}{Rev. Mod. Phys. {\bf 77}, 259 (2005).}
%%%%%%%%%%%%%%%%%%%%%%%%%%%%%%%%%%%%%%%%%%%%%%%%%%%%%%
%%%%%%%%%%%%%%%%%%%%%%%%%%%%%%%%%%%%%%%%%%%%%%%%%%%%%
\bibitem{Uli2} U. Schollw\"ock,
{\it The density-matrix renormalization group in the age of matrix product states}, \href{https://www.sciencedirect.com/science/article/pii/S0003491610001752}{Ann. Phys. {\bf 326}, 96 (2011).}
%%%%%%%%%%%%%%%%%%%%%%%%%%%%%%%%%%%%%%%%%%%%%%%%%%%%%%
%%%%%%%%%%%%%%%%%%%%%%%%%%%%%%%%%%%%%%%%%%%%%%%%%%%%%
\bibitem{banuls} M. C. Ba\~nuls,
{\it Tensor Network Algorithms: A Route Map}, \href{https://www.annualreviews.org/doi/full/10.1146/annurev-conmatphys-040721-022705}{Annual Review of Cond. Matt. Phys. {\bf 14}, 173 (2023).}
%%%%%%%%%%%%%%%%%%%%%%%%%%%%%%%%%%%%%%%%%%%%%%%%%%%%%%
%%%%%%%%%%%%%%%%%%%%%%%%%%%%%%%%%%%%%%%%%%%%%%%%%%%%%
\bibitem{Com}
The series of numbers represent the average occupations.
%%%%%%%%%%%%%%%%%%%%%%%%%%%%%%%%%%%%%%%%%%%%%%%%%%%%%%
%%%%%%%%%%%%%%%%%%%%%%%%%%%%%%%%%%%%%%%%%%%%%%%%%%%%%
\bibitem{bak}
P. Bak,
{\it Commensurate phases, incommensurate phases and the
devil's staircase},
\href{https://iopscience.iop.org/article/10.1088/0034-4885/45/6/001/meta}{Rep. Prog. Phys. {\bf 45}, 587 (1982)}.
%%%%%%%%%%%%%%%%%%%%%%%%%%%%%%%%%%%%%%%%%%%%%%%%%%%%%%
%%%%%%%%%%%%%%%%%%%%%%%%%%%%%%%%%%%%%%%%%%%%%%%%%%%%%
\bibitem{lockin}
See also K. Inagaki, K. Nakatsugawa, and S. Tanda,
{\it Lock-in transition of the charge-density waves of
quasi-one-dimensional conductors: origin of the first-order
transition},
\href{https://arxiv.org/abs/2304.11525}{arXiv:2304.11525}.
%%%%%%%%%%%%%%%%%%%%%%%%%%%%%%%%%%%%%%%%%%%%%%%%%%%%%%
%%%%%%%%%%%%%%%%%%%%%%%%%%%%%%%%%%%%%%%%%%%%%%%%%%%%%
\bibitem{Cho}
D. Cho, G. Gye, J. Lee, S-H. Lee, L. Wang, S-W. Cheong, and H. W. Yeom,
{\it Correlated electronic states at domain walls of a Mott-charge-density-wave insulator $1T-TaS_2$,}
\href{https://doi.org/10.1038/s41467-017-00438-2}{Nat. Comm. {\bf 8}, 392 (2017)}.
%%%%%%%%%%%%%%%%%%%%%%%%%%%%%%%%%%%%%%%%%%%%%%%%%%%%%%
%%%%%%%%%%%%%%%%%%%%%%%%%%%%%%%%%%%%%%%%%%%%%%%%%%%%%
\bibitem{nakamura}
M. Nakamura,
{\it Tricritical behavior in the extended Hubbard chains,}
\href{https://doi.org/10.1103/PhysRevB.61.16377}{Phys. Rev. B {\bf 61}, 16377 (2000)}.
%%%%%%%%%%%%%%%%%%%%%%%%%%%%%%%%%%%%%%%%%%%%%%%%%%%%%%
%%%%%%%%%%%%%%%%%%%%%%%%%%%%%%%%%%%%%%%%%%%%%%%%%%%%%
\bibitem{SM}
See Supplemental Material at [URL will be inserted by publisher] for details.
%%%%%%%%%%%%%%%%%%%%%%%%%%%%%%%%%%%%%%%%%%%%%%%%%%%%%%%
%%%%%%%%%%%%%%%%%%%%%%%%%%%%%%%%%%%%%%%%%%%%%%%%%%%%%%

\bibitem{moessner} M. Schulz, C. A. Hooley, R. Moessner, and F. Pollmann, {\it Stark Many-Body Localization},
\href{https://doi.org/10.1103/PhysRevLett.122.040606}{
Phys. Rev. Lett. {\bf 122}, 040606 (2019).}

\end{thebibliography}
\end{document}